\documentstyle[12pt,epsfig]{article}

\def\be{\begin{equation}}
\def\ee{\end{equation}}
\def\bea{\begin{eqnarray}}
\def\eea{\end{eqnarray}}
\def\nnb{\nonumber}

\begin{document}

\hfill{SNUTP 97-105} 

\vskip .4in
\begin{center}
{\LARGE {\bf Radiative Muon Capture \\
in Heavy Baryon Chiral Perturbation Theory }}
\vskip .4in
{\large Shung-ichi Ando\footnote{E-mail: ando@zoo.snu.ac.kr}
 and Dong-Pil Min\footnote{E-mail: dpmin@phya.snu.ac.kr}}\\[.1in]
{\large \it Department of Physics and Center for Theoretical Physics,}\\
{\large \it Seoul National University, Seoul 151-742, Korea}\\

\vskip 2cm

\begin{quotation}
The radiative muon capture(RMC) on a proton is
analyzed by means of heavy baryon chiral perturbation theory. The emitted 
photon energy spectrum is calculated and compared with the experimental
data by taking the spin sum on the muonic atom states. 
We find that one-loop order corrections to the tree order amplitude
modify the photon spectrum by 
less than five 
percent. This calculation
supports that the theory is under a quantitative control as far as the chiral
perturbation expansion is concerned and indicates that the discrepancy between
the pseudo-scalar coupling constant required by the RMC experiment and  the
one deduced from ordinary muon capture, the value of which is also supported
by chiral perturbation calculations, will remain unexplained from the 
theoretical side. 

\vskip 4mm

\noindent
Key wards; $\mu^-+p \rightarrow \nu n\gamma$, heavy baryon chiral perturbation 
theory, pseoudo-scalar coupling

\end{quotation}
\end{center}

\newpage

The {\it induced} pseudo-scalar coupling constant $g_P$ 
was determined from the ordinary muon capture (OMC) reaction
on a proton $(p+\mu\rightarrow n+\nu)$\cite{bardin} to be $g_P=8.7\pm 1.9$.
Despite its large error bar, this value is clearly consistent 
with the theoretical prediction by 
Bernard {\it et al.}~\cite{bernard} 
using heavy baryon chiral perturbation theory (HBChPT),
$g_P(q^2=-0.88m_\mu^2)=8.44\pm 0.16$. This is also comparable to
the value that Fearing {\it et al.}~\cite{fearing0}
evaluated by means of HBChPT in their work on OMC, 
$g_P(-0.88m_\mu^2)=8.21\pm0.09$. 
Together with
the PCAC prediction $g_P^{PCAC}(-0.88m_\mu^2)=8.42$, all theoretical
investigations agree with the experimental result on OMC. 
However, the momentum transfer involved in OMC is far
from the pion pole, $q^2=m_\pi ^2$, 
where the pseudo-scalar coupling should play
important role in the reaction amplitude, accounting for the
large error bar.

Since RMC involves a momentum transfer closer to the pion mass, it is
considered to be more suitable to measure the constant
$g_P$ and lower the error bar. For this purpose, 
the photon energy spectrum from the radiative muon capture (RMC)
($p+\mu\rightarrow n+\nu+\gamma$) has been measured in TRIUMF   \cite{jonkmans}
and compared to the model prediction of \cite{fearing1}.
Surprisingly, the experimentally detected photon spectrum could be explained
only if the pseudo-scalar coupling constant is enhanced in the model
by a factor of 1.5 relative to the value given by PCAC \cite{opat} or that
determined in OMC. A calculation to tree order, recently reported by Meissner
{\it et al.}\cite{kubodera}, further confirms this discrepancy.

The purpose of this paper is to see whether or not this discrepancy can
be eliminated by higher order terms in the treatment of the strong interaction
sector of the process.   
It is natural to ask whether any important
Feynman diagrams have been ignored in the phenomenological
model, in particular in light of the direct chiral perturbation calculation
 of the pseudo-scalar coupling constant by Bernard {\it et al.}~\cite{bernard}
which agrees with the PCAC prediction. Experiments are currently
being planned\cite{hasinoff} to increase the precision. 

In this work, we shall calculate the RMC amplitude and 
the photon energy spectrum using the HBChPT up 
to the next-to-next to the leading order (${\rm N^2LO}$), that is, to 
one-loop order.  We shall also investigate 
the photon energy spectrum by taking various ans\"atze on the spin states
of the muonic atom.

Heavy baryon chiral perturbation theory(HBChPT)\cite{georgi} provides a 
systematic way of making, in the presence of nucleons, a chiral 
perturbation expansion 
in powers of $Q/\Lambda_\chi$, where $Q$ is a typical momentum scale
and/or the pion mass and 
$\Lambda_\chi$ is the chiral symmetry scale, 
$\Lambda_\chi\sim 1 {\rm GeV}$.
Since the momentum transfer involved in RMC can be of the order of 
the muon mass, $m_\mu \sim 0.1 \Lambda_\chi$, the chiral expansion
is expected to converge sufficiently rapidly.

The Feynman graphs contributing to RMC can be classified into two
classes as shown in Fig. 1:
\begin{figure}[h]
\setcounter{figure}{0}
\renewcommand\thefigure{\arabic{figure}}
\begin{center}
\epsfig{file=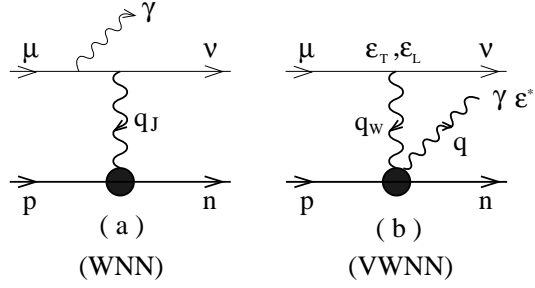,width=7cm }
\caption{Diagrams for three and four point Greens functions. }
\end{center}
\end{figure}
(a) the first corresponds to those graphs where the photon line is
attached to the lepton, therefore, leaving the nucleon line to form
the 3-point vertex of $WNN$ (weak current-nucleon-nucleon), 
(b) the second corresponds to the graphs where the photon is
attached to the nucleon line and the vertex with the exchanged
pion coupled to the weak boson, 
which is schematically a 4-point 
vertex of $VWNN$ 
(electro-magnetic current-weak current-nucleon-nucleon). 
Indeed, those two vertex graphs shown as blobs in Fig. 1 
can be expressed in terms of Green's functions as follows,
\bea
J^{f}_\beta(q_J,k) &=&
    - \chi^\dagger_n S_N^{-1}(k')\prod^3_{i=1}\int\frac{d^4x_i}{(2\pi)^4}
    {\rm exp}\left[-i(q_J\cdot x_1-k'\cdot x_2+k\cdot x_3)\right] \nnb \\
& & \times \langle 0|T\ W^f_\beta(x_1)N(x_2)\bar{N}(x_3)
    |0\rangle
    S_N^{-1}(k)\chi_p,
 \label{nucleon_current_generating} \\
M^{ef}_{\alpha\beta}(q,q_W,k) &=& -i\chi^\dagger_n S_N^{-1}(k') \prod^4_{i=1}
    \int\frac{d^4x_i}{(2\pi)^4}{\rm exp}\left[-i(-q\cdot x_1 + q_W\cdot x_2
    - k'\cdot x_3 + k\cdot x_4)\right] \nnb \\
& & \times
    \langle 0|T\  V^e_\alpha(x_1)W^f_\beta(x_2)N(x_3)
    \bar{N}(x_4) |0\rangle
     S_N^{-1}(k) \chi_p,
 \label{nucleon_tensor_generating}
\eea
where $e$,$f$ and $\alpha$,$\beta$ are the iso-spin and Lorentz indices, 
respectively, and $T$ stands for the time ordering on the currents and
fields appearing on its right. Note that the vacuum expectation
value in the above equations is the Green's function.
The four-momenta carried by proton,
neutron, neutrino, muon are  
denoted by the particle symbols,  $p$, $n$, $\nu$, $\mu$, respectively,
whereas the four-momentum of the photon by $q$. 
The momentum transfers described in  Fig. 1 can be
written as $q_W=\mu-\nu$ and $q_J=q_W-q$. Here
$\chi_p$ ($\chi_n$) is a 
two-component spinor of the proton (neutron) 
with the normalization condition\footnote{Actually $\chi$ depends on $v^\mu$.
Here we have set $v^\mu=(1,\vec{0})$.} 
$\chi^\dagger\chi=E+m_N$, and 
$S_N^{-1}(k)$ is the inverse of the {\it heavy} nucleon propagator 
to be derived below.
In the above, $k$ ($k'$) denotes 
the residual momentum of the proton (neutron) so that
\bea
p^\mu &=& m_N v^\mu + k^\mu , \\
n^\mu &=& m_N v^\mu + k'^\mu , 
\eea
with the velocity four-vector $v_\mu = (1, \vec{0})$. 
$m_N$ is the {\it physical} nucleon mass.
The {\it heavy} nucleon field $N$ is defined from the nucleon field 
$\Psi_N$ as
\be
N=e^{im_Nv\cdot x}P_+\Psi_N,
\ee
where $P_+$ is the projection operator defined by 
$P_+=\frac{1}{2}(1+v\cdot \gamma)$. 

The Green's functions in Eqs.(\ref{nucleon_current_generating}) and
(\ref{nucleon_tensor_generating}) are to be obtained from the effective 
chiral lagrangian with nucleons and pions, the expression of which reads
\be
{\cal L} = {\cal L}_0+{\cal L}_1+{\cal L}_2+\cdots,
 \label{lagrangians}
\ee
where ${\cal L}_0$ is the leading order lagrangian given in \cite{tsp1}
and 
${\cal L}_1$ is of the $1/m_N$ correction (NLO) 
which will be specified below.
${\cal L}_2$ is the next-to-next leading order (${\rm N^2LO}$)
effective lagrangian. 
The ellipses stand for higher order lagrangians irrelevant
for our calculation.
${\cal L}_1$ reads\cite{tsp1} 
\bea
{\cal L}_1 &=&  \frac{1}{2m_N} \bar{N}
    \left[ - D^2 + (v\cdot D)^2  + 2 g_A \{ v\cdot \Delta , S\cdot D\}
    \right. \nnb \\
& & + b_1 {\rm Tr}(\chi_+)  + b_2(\chi_+- \frac{1}{2}{\rm Tr}(\chi_+))
    + (g_A^2 + b_3) (v\cdot \Delta)^2 + b_4 \Delta\cdot\Delta \nnb \\
& & \left. - [S^\mu,S^\nu] \left( 2(1+b_5) \Delta_\mu \Delta_\nu
    + i(1+b_6) f^+_{\mu\nu} + i(1+b_7) v^S_{\mu\nu}
    \right) \right] N ,
 \label{lagrangian1}
\eea
where we have adopted the notations 
$D_\mu$, $\Delta_\mu$, $\chi_+$, $f^+_{\mu\nu}$ 
and $v^S_{\mu\nu}$ as defined in Ref.\cite{tsp1}, except for 
an additional multiplication factor 1/2 for our $f^+_{\mu\nu}$.  
$g_A$ is the axial-vector current coupling constant, and $b_i$ are
the {\it low energy constants} that cannot be fixed by the theory,
but will be determined from experiments. We should note that
for RMC, only the two constants $b_6$ and $b_7$ are relevant. 
They are determined from 
the anomalous magnetic moments of nucleon 
as $b_6=\kappa_V=3.71$
\footnote{Up to $N^2LO$, $\kappa_V$ is renormalized by
$b_6 -\frac{g_A^2m_\pi m_N}{4\pi f_\pi^2}=\kappa_V$.},
 $b_7=\kappa_S=-0.12$, where $\kappa_V$ and $\kappa_S$ are
the iso-vector and iso-scalar anomalous magnetic moment, respectively.
The ${\rm N^2LO}$ lagrangian ${\cal L}_2$
containing low energy constants and an anomaly term reads 
\begin{eqnarray}
\lefteqn{{\cal L}_2 =
 i\alpha _9^{(2)}{\rm Tr}(L_{\mu \nu }\nabla ^\mu U\nabla ^\nu U^{\dagger }
 +R_{\mu \nu }\nabla ^\nu U^{\dagger }\nabla ^\nu U)
 +\alpha _{10}^{(2)}{\rm Tr}(L_{\mu \nu }UR_{\mu \nu }U^{\dagger })
 +{\cal L}_{WZ}}
 \nnb \\
&& +\frac{1}{4m_N^2}\bar{N}\left[
 iD^\alpha v\cdot DD_\alpha
 - 2i[S^\alpha,S^\beta]D_\alpha v\cdot D D_\beta
 + i g_A [ 2 D\cdot \Delta S\cdot D - 2 D^\alpha S\cdot\Delta D_\alpha
 \right.\nnb \\
& & + 2 S\cdot D \Delta\cdot D + 2 v\cdot D S\cdot \Delta v\cdot D
 -2\{S\cdot D,\{v\cdot D,v\cdot \Delta\}\}
     -i\epsilon^{abcd}D_a\Delta_b D_cv_d]
\nnb\\
&& + i[-v^\alpha D^\beta + v^\beta D^\alpha
    - 2 v^\alpha[S^\beta,S\cdot D] + 2 v^\beta[S^\alpha,S\cdot D]
 +2g_A(v^\alpha S^\beta-v^\beta S^\alpha)v\cdot \Delta]B_{\alpha\beta}
   \nnb \\
& & \left.
 + iB_{\alpha\beta}[v^\alpha D^\beta-v^\beta D^\alpha
    - 2 v^\alpha[S^\beta,S\cdot D] + 2 v^\beta [S^\alpha,S\cdot D]
    - 2g_A(v^\alpha S^\beta - v^\beta S^\alpha)v\cdot \Delta]
 \right] N \nnb \\
&& + \frac 1{(4\pi f_\pi) ^2}
 \bar{N}\left[ c_3v^\alpha [D^\beta ,f_{\alpha \beta }^{+}]
 +c_4[S^\alpha ,S^\beta ]\{v\cdot D ,f_{\alpha \beta }^{+}\}
 +c_8g_Av^\alpha S^\beta [v\cdot \Delta ,f_{\alpha \beta }^{+}]\right. 
\nonumber \\
&&\left. +c_{12}[S^\alpha ,S^\beta ]\{v\cdot D,v_{\alpha ,\beta }^S\}
 +c_{13}g_AS^\alpha [D_\alpha ,f_{\alpha \beta }^{-}]
 +ic_{14}g_AS^\alpha [D_\alpha ,\chi _{-}]\right] N ,
\end{eqnarray}
with 
\be
B_{\alpha\beta}=i\frac{b_6}{2}f^+_{\alpha\beta}
 +i\frac{b_7}{2}v^S_{\alpha\beta}
\ee
where ${\cal L}_{WZ}$ is the Wess-Zumino lagrangian\cite{wzlagrangian}.
Note that we have eight low-energy constants, among which
seven of them are needed for this work:
$\alpha_9^{(2)}+\alpha_{10}^{(2)}=1.43\times 10^{-3}$ determined
from a rare pion process\cite{donoghue},
$c_3=5.34$ ($c_{13}=2.37$) from the iso-vector vector 
(axial-vector) radius,
$c_{14}=-1.37$ from the Goldberger-Treiman discrepancy\cite{bernard,fearing0}, 
$c_8=-3.27$ from the $\Delta$ contribution to $E^{(-)}_{0+}$\cite{bernardE},
$c_4=-22.27$ and $c_{12}=-0.79$ 
from the $\rho$, $\omega$ and $\Delta$
contributions to $P^{0,-}_3$\cite{bernardP}. 
{\it In short, the constants are completely determined for calculating up to 
the ${\rm N^2LO}$ order in chiral perturbation expansion.}

We are now in a position to calculate the relevant Feynman graphs.
The chiral power counting rule for A-nucleon processes is that for a Feynman
graph with $V_i$ vertices of type $i$, $L$ loops, and $C$ separately connected
pieces, the power index of $Q$  is given by
$\eta=4-A-2C+2L+\sum_i V_i \Delta_i$ with $\Delta_i = d_i+n_i/2 -2$,
where $n_i$ is the number of nucleon lines and $d_i$ is the number of 
derivatives or powers of $m_\pi$ at the $i-$type vertex. In the presence 
of an external gauge field, $\Delta_i $ is constrained by chiral symmetry
to be 
$\Delta_i \ge -1$\cite{rho}. Thus
the leading order of matrix elements $J$ is $O(1)$ and that of $M$ 
is $O(Q^{-1})$. 
However, the leading order amplitudes of Fig. 1(a) and Fig. 1(b) are of the
same chiral order, because the muon propagator in  Fig. 1(a) is
of order $Q^{-1}$ since it carries 
the photon momentum in the denominator.

Now we split the weak-current into the $V-A$ form 
so that the $J$ and $M$ can be written
\begin{eqnarray}
J &=& J_{V} - J_{A} , \\
M &=& M_{V} - M_{A} ,
\end{eqnarray}
The most general forms of $J$ turn out to be, (with $v_\mu = (1, \vec{0})$)
\be
\left.
\begin{array}{c}
 J^0_V(q_J,k=0)= f_1^V, \ \
 \vec{J}_V(q_J,k=0)=i\,\vec{\sigma}\times\hat{q}_J f_2^V+\hat{q}_J f_3^V,  \\
 J^0_A(q_J,k=0)= \vec{\sigma}\cdot \hat{q}_J f_3^A, \ \
 \vec{J}_A(q_J,k=0)=\vec{\sigma}f_1^A+\hat{q}_J\vec{\sigma}\cdot
 \hat{q}_J f_2^A , 
\end{array}
\right.
\label{ffja}
\ee
where as is done in what follows, the initial and final state nucleon 
spinors are omitted.
$f^V_i$ and $f^A_i$ denote the nucleon vector and
axial-vector form factors, respectively. 
They read as, up to ${\rm N^2LO}$,
\bea
f_1^V &=& 1 +\frac{c_3}{(4\pi f_\pi)^2} q_J^2 
 -\frac{1+17g_A^2}{18(4\pi f_\pi)^2}q_J^2
 +\frac{1}{(4\pi f_\pi)^2}
  \left[\frac{2}{3}(1+2g_A^2)m_\pi^2 -\frac{1+5g_A^2}{6}q_J^2\right]f_0(q_J) 
 \nnb \\
&& + \frac{1}{4m_N^2}(-\frac 3 2 +\kappa_V) q_J^2,
 \label{f_form;f1v} \\
f_2^V &=& \frac{1}{2m_N}(1+\kappa_V)|\vec{q}_J|
 +\frac{g_A^2}{64\pi f_\pi^2 m_\pi}q_J^2|\vec{q}_J|
 +\left(\frac{g_A}{4\pi f_\pi}\right)^2\frac{\pi(4m_\pi^2-q_J^2)}{4m_\pi}
 m_0(\vec{q}_J)|\vec{q}_J| ,
 \label{f_form;f2v} \\
f_3^V &=& \frac{1}{2m_N}|\vec{q}_J|,
 \label{f_form;f3v} \\
f_1^A &=& g_A\left[1+\left(\frac{c_{13}}{(4\pi f_\pi)^2}-\frac{1}{8m_N^2}
 \right)q_J^2\right], 
 \label{f_form;f1a} \\
f_2^A &=& g_A\left[\frac{c_{13}}{(4\pi f_\pi)^2}
 +\Delta_\pi(q_J)\left(1-\frac{2m_\pi^2c_{14}}{(4\pi f_\pi)^2}
 +\frac{1}{8m_N^2}q_J^2\right)\right]|\vec{q}_J|^2,
 \label{f_form;f2a} \\
f_3^A &=& \frac{g_A}{2m_N}[1+\Delta_\pi(q_J)q_J^2]|\vec{q}_J|,
 \label{f_form;f3a} 
\eea 
with
\bea
f_0(q) &=& \int^1_0dx {\rm ln}[1-x(1-x)\frac{q^2}{m_\pi^2}], \\
m_0(\vec{q}) &=& 1-\int^1_0dx\frac{1}{\sqrt{1+x(1-x)\vec{q}^2/m_\pi^2}}.
\eea
The common factor $2m_N$ is omitted in the above equations, and 
$g_A=1.25$. 
The convergence of the form factors $f_i^{V(A)}$ is found to
be quite good as is discussed in  Ref.\cite{bernard,fearing0,tsp1}.

Under the Coulomb gauge,
the renormalized {\it inverse} propagators for our calculation can be written  simply
as
\bea
S_N^{-1}(k) &=&  v\cdot k
 + \frac{1}{2 m_N} [ k^2 - v\cdot k^2 ] 
, \\
\Delta_\pi^{-1}(q) &=& q^2 - m_\pi^2 .
\eea

We choose the coordinate frame such that 
the neutrino lies in the $z$-direction and
the photon in the $x$-$z$ plane, respectively, i.e., $\hat{\nu}=(0,0,1)$ and 
$\hat{q}=({\rm sin}\theta,0,{\rm cos}\theta)$, where
$\theta$ is the angle between the neutrino and the photon.
Then we can decompose $M$ into so-called {\it reduced} amplitudes 
for each muon spin states lying along $z$-axis, $ m_s =\pm 1/2$ \cite{hemmert},
\bea
M &=& - \vec{\epsilon}_T\cdot \vec{M}_T \ \delta_{m_s,1/2} 
+ \epsilon_{L,3}\  M_L \ \delta_{m_s,-1/2},
\eea
where
$ \epsilon^{\beta}_T =\epsilon^{\beta}_{(1/2)}$
and
$\epsilon^{\beta}_{L} = \epsilon^{\beta}_{(-1/2)}$
with
$\epsilon^{\beta}_{(m_s)}
\equiv 
\bar{u}_\nu\gamma_5(1-\gamma^\beta)u_\mu^{(m_s)}$.

Then generally one can decompose them into different {\it spin} 
operators as 
specified in Tables 1 and 2,
\bea
& & - \vec{\epsilon}_T\cdot \vec{M}_{A,T} = 
 \sum_{i=1}^{6} A_i {\cal O}_{a,i} ,
\ \ \ 
M_{A,L} = 
 \sum_{i=7}^{10} A_i {\cal O}_{a,i}  ,
 \label{ffva2} \\
& & - \vec{\epsilon}_T\cdot \vec{M}_{V,T} = 
 \sum_{i=1}^{9} B_i {\cal O}_{b,i}  ,
\ \ \ 
M_{V,L} = 
 \sum_{i=10}^{13} B_i {\cal O}_{b,i} ,
 \label{ffvv2} 
\eea
where ${\cal O}_{a,i}$ and $A_i$ (${\cal O}_{b,i}$ and $B_i$) are operators 
and corresponding form factors, respectively.
The Ward-Takahashi (WT) identity was found to be quite useful in reducing
redundancy in the form factors.
Summation runs over all possible effective operators.
\begin{table}[h]
\begin{center}
\begin{tabular}{|c||c|c|c|c|} \hline
$i$ & ${\cal O}_{b,i}$ & $A_i^{LO}$ & $A_i^{NLO}$ &
 $|m_\mu A_i^{loop}|_{max}$ \\ \hline \hline
$1$ & $i\hat{q}\cdot \vec{\epsilon}^*\times \vec{\epsilon}_T$ & $0$ &
 $- \frac{g_A}{2m_N}(1+\kappa_V)$ &  $0.030$\\ \hline
$2$ & $\vec{\sigma} \cdot \vec{\epsilon}^*\ \vec{\epsilon}_T\cdot\hat{q}$ &
 $0$ & $- \frac{g_A}{2m_N}(1+\kappa_S)$ & $0.207$  \\ \hline
$3$ & $\vec{\sigma}\cdot\hat{q}\ \vec{\epsilon}^*\cdot\vec{\epsilon}_T$ &
 $- g_A \Delta_\pi(q_J) E_\gamma$ & $\frac{g_A}{2m_N}(1+\kappa_S)$ & 
 $0.232$ \\ \hline
$4$ & $\vec{\sigma}\cdot \hat{q}_W\ \vec{\epsilon}^*\cdot\vec{\epsilon}_T$ &
 $g_A \Delta_\pi(q_J) E_\nu$ & $0$ & 
 $0.001$ \\ \hline 
$5$ & $\vec{\sigma}\cdot \hat{q}\ 
 \vec{\epsilon}^*\cdot \hat{q}_W\  \vec{\epsilon}_T\cdot \hat{q}$ & $0$
 & $0$ & $0.001$ \\ \hline
$6$ & $\vec{\sigma}\cdot \hat{q}_W\ 
 \vec{\epsilon}^*\cdot \hat{q}_W\  \vec{\epsilon}_T\cdot \hat{q}$ & $0$
 & $0$ & $0.001$ \\ \hline  \hline
$7$ & $i \vec{\epsilon}^*\cdot \hat{q}\times \hat{q}_W$ &
 $0$ & $ \frac{-g_A}{2m_N}(1+\kappa_V)$
 & $0.033$ \\ 
 & & & $\times(1+\Delta_\pi(q_W)m_\mu E_\nu)$ &  $$
 \\ \hline
$8$ & $\vec{\sigma}\cdot \vec{\epsilon}^*$ & 
 $- g_A \Delta_\pi(q_W) m_\mu$ & $\frac{-g_A}{2m_N}\left[
 (1+\kappa_S) \right.$ & $0.219$ \\
 & & & $\times(1+\Delta_\pi(q_W)m_\mu E_\nu)y$ & $$ \\ 
 & & & $\left.-(1-\Delta_\pi(q_W)m_\mu E_\gamma) \right]$ &
 $$ \\ \hline
$9$ & $\vec{\sigma}\cdot \hat{q}\  \vec{\epsilon}^*\cdot \hat{q}_W$ &
 $g_A \Delta_\pi(q_J)E_\gamma(1+$ & 
 $\frac{-g_A}{2m_N}(1+\kappa_S)$ &  $0.184$ \\   
 & & $2\Delta_\pi(q_W)m_\mu E_\nu)$ & 
 $\times( 1+\Delta_\pi(q_J)m_\mu E_\nu)$ & $$
 \\ \hline
$10$ & $\vec{\sigma}\cdot \hat{q}_W\ \vec{\epsilon}^*\cdot \hat{q}_W$ &
 $-g_A\Delta_\pi(q_J)E_\nu(1+$ & $0$ & $0.015 $  \\
 & & $2\Delta_\pi(q_W)m_\mu E_\nu)$ &  & $$
 \\ \hline
\end{tabular}
\caption{Operators and form factor $A_i$ for the $M_A$:
where $E_\gamma$ ($E_\nu$) is the photon (neutrino) energy. 
And $y={\rm cos}\theta$, $\theta$ is the angle between the neutrino 
and photon.}
\end{center}
\end{table}
\begin{table}[h]
\begin{center}
\begin{tabular}{|c||c|c|c|} \hline
$i$ & ${\cal O}_{a,i}$ & $B_i^{NLO}$ & $|m_\mu B_i^{loop}|_{max}$ 
 \\ \hline \hline
$1$ & $\vec{\epsilon}^*\cdot \vec{\epsilon}_T$ & 
 $- \frac{1}{2m_N}$ & $0.080$ \\ \hline
$2$ & $\vec{\epsilon}^*\cdot  \hat{q}_W\  \vec{\epsilon}_T\cdot \hat{q}$ &
 $0$ & $\sim 10^{-4}$\\ \hline 
$3$ & $i \vec{\sigma}\cdot (\vec{\epsilon}^{*}\times\vec{\epsilon}_T)$ &
 $\frac{1}{2m_N}(1+\kappa_V)$ & $0.007$ \\ \hline
$4$ & $i \vec{\sigma}\cdot (\hat{q} \times\hat{q}_W) \  \vec{\epsilon}^{*}
 \cdot \vec{\epsilon}_T$ & $0$ & $0.003$ \\ \hline
$5$ &$i\vec{\sigma}\cdot(\vec{\epsilon}^*\times\hat{q}_W)\ \vec{\epsilon}_T
 \cdot \hat{q}$ & $0$ & $0.003$ \\ \hline
$6$ &$i\vec{\sigma}\cdot (\vec{\epsilon}^*\times \hat{q})\ \vec{\epsilon}_T
 \cdot \hat{q}$ & $0$ & $0.009$ \\ \hline 
$7$ &$-i\vec{\sigma}\cdot(\vec{\epsilon}_T\times\hat{q})\  \vec{\epsilon}^*
 \cdot \hat{q}_W$ & $0$ & $0.003$ \\ \hline
$8$ & $-i\vec{\sigma}\cdot(\vec{\epsilon}_T\times\hat{q}_N)\ 
 \vec{\epsilon}^* \cdot \hat{q}_W$ & $0$ & $0.010$ \\ \hline
$9$ & $i\vec{\sigma}\cdot (\hat{q}\times \hat{q}_W)
 \vec{\epsilon}^*\cdot \hat{q}_W \vec{\epsilon}_T\cdot\hat{q}$ & $0$ & 
 $\sim 10^{-4}$ \\ \hline \hline
$10$ & $\vec{\epsilon}^* \cdot \hat{q}_W$ & $\frac{1}{2m_N}$ & $0.029$ 
   \\ \hline  
$11$ & $i\vec{\sigma}\cdot (\hat{q} \times \hat{q}_W)\  \vec{\epsilon}^*\cdot
 \hat{q}$ & $0$ & $0.020$ \\ \hline
$12$ & $i\vec{\sigma}\cdot \vec{\epsilon}^*\times \hat{q}_W$ & 
 $- \frac{1}{2m_N}(1+\kappa_V)$ & $0.014$   \\ \hline
$13$ & $i \vec{\sigma}\cdot \vec{\epsilon}^*\times \hat{q}$ &
 $- \frac{1}{2m_N}(1+\kappa_V)$ & $0.010$ \\ \hline  
\end{tabular}
\label{table;A}
\caption{Operators and form factor $A_i$ for the $M_V$:
Operators are identical to those occurred in Ref.\protect\cite{hemmert}}
\end{center}
\end{table}

Some LO and NLO contributions in $A_i$ in Table 1 contain the pion 
propagator taken at $q_W$,  which make the difference between RMC
and OMC.
We present the results in Tab. 1 and 2 calculated in Coulomb gauge.
The formulae for the
matrix elements are quite lengthy and uninstructive; we leave their
explicit expressions to a forthcoming paper\cite{sa}. 
For the contribution of ${\rm N^2LO}$ 
we give the values of their  maximum among the entire range of photon
energies.
Among the contributions of this order, the most
important one comes from the intermediate excitation of a 
$\Delta$ contributing to  the term proportional to $c_4$. We have multiplied
by $m_\mu$ in the last column of Table 1 to make the numbers dimensionless.
One can see that at their maximum, some of them  are comparable with 
the $1/m_N$ corrections. 
However, in the total spectrum,
their correction is 
less than five 
percent as is shown in
Fig. 2.

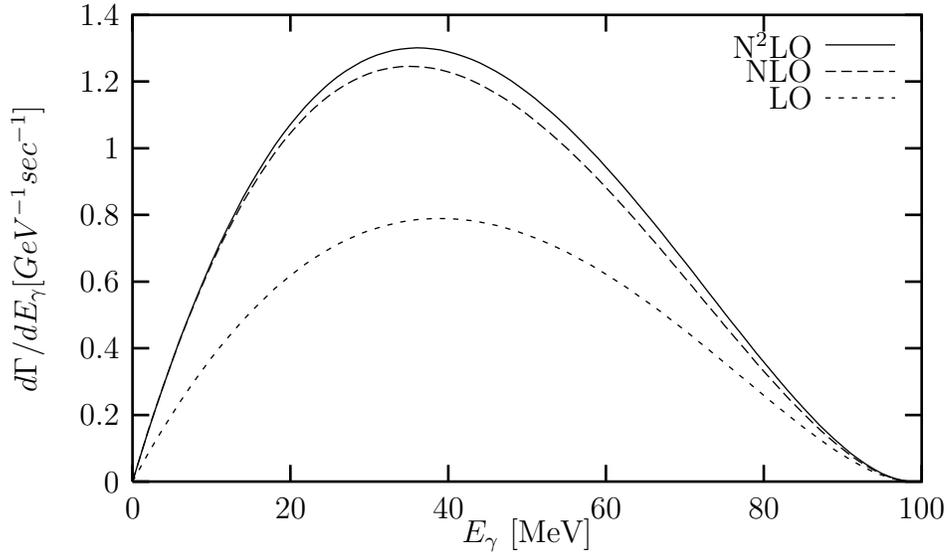
\begin{figure}[p]
\begin{center}
\setlength{\unitlength}{0.1bp}
\begin{picture}(3600,2160)(0,0)
\put(3024,1747){\makebox(0,0)[r]{LO}}
\put(3024,1847){\makebox(0,0)[r]{NLO}}
\put(3024,1947){\makebox(0,0)[r]{${\rm N^2LO}$}}
\put(1950,100){\makebox(0,0){$E_\gamma$ [MeV]}}
\put(100,1180){%
\makebox(0,0)[b]{\shortstack{$d\Gamma/dE_\gamma[GeV^{-1}sec^{-1}]$}}%
}
\put(3437,200){\makebox(0,0){100}}
\put(2842,200){\makebox(0,0){80}}
\put(2247,200){\makebox(0,0){60}}
\put(1653,200){\makebox(0,0){40}}
\put(1058,200){\makebox(0,0){20}}
\put(463,200){\makebox(0,0){0}}
\put(413,2060){\makebox(0,0)[r]{1.4}}
\put(413,1809){\makebox(0,0)[r]{1.2}}
\put(413,1557){\makebox(0,0)[r]{1}}
\put(413,1306){\makebox(0,0)[r]{0.8}}
\put(413,1054){\makebox(0,0)[r]{0.6}}
\put(413,803){\makebox(0,0)[r]{0.4}}
\put(413,551){\makebox(0,0)[r]{0.2}}
\put(413,300){\makebox(0,0)[r]{0}}
\end{picture}
\caption{Photon spectrum of contributions of each order. }
\end{center}
\end{figure}

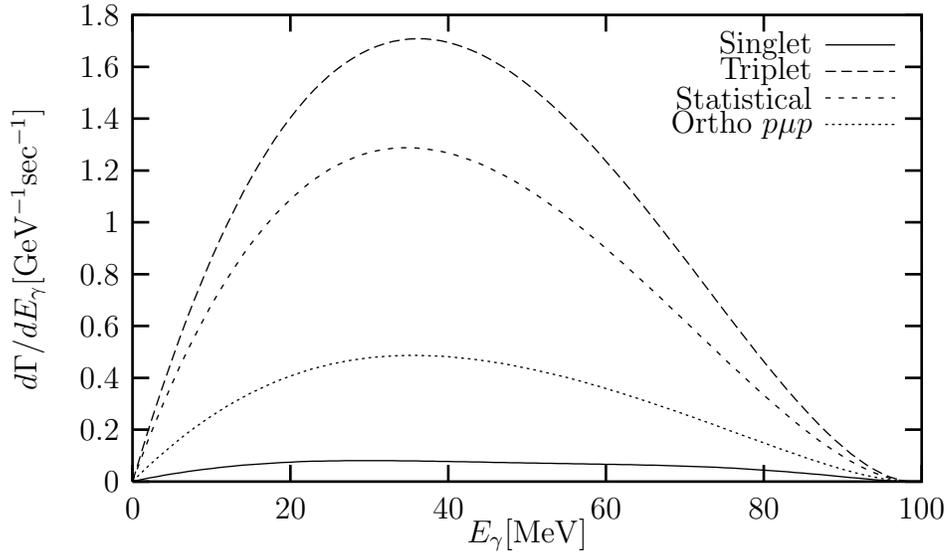
\begin{figure}
\begin{center}
\setlength{\unitlength}{0.1bp}
\begin{picture}(3600,2160)(0,0)
\put(3024,1647){\makebox(0,0)[r]{Ortho $p\mu p$}}
\put(3024,1747){\makebox(0,0)[r]{Statistical}}
\put(3024,1847){\makebox(0,0)[r]{Triplet}}
\put(3024,1947){\makebox(0,0)[r]{Singlet}}
\put(1950,100){\makebox(0,0){$E_\gamma [{\rm MeV}]$}}
\put(100,1180){%
\makebox(0,0)[b]{\shortstack{$d\Gamma/dE_\gamma [{\rm GeV^{-1}sec^{-1}}]$}}%
}
\put(3437,200){\makebox(0,0){100}}
\put(2842,200){\makebox(0,0){80}}
\put(2247,200){\makebox(0,0){60}}
\put(1653,200){\makebox(0,0){40}}
\put(1058,200){\makebox(0,0){20}}
\put(463,200){\makebox(0,0){0}}
\put(413,2060){\makebox(0,0)[r]{1.8}}
\put(413,1864){\makebox(0,0)[r]{1.6}}
\put(413,1669){\makebox(0,0)[r]{1.4}}
\put(413,1473){\makebox(0,0)[r]{1.2}}
\put(413,1278){\makebox(0,0)[r]{1}}
\put(413,1082){\makebox(0,0)[r]{0.8}}
\put(413,887){\makebox(0,0)[r]{0.6}}
\put(413,691){\makebox(0,0)[r]{0.4}}
\put(413,496){\makebox(0,0)[r]{0.2}}
\put(413,300){\makebox(0,0)[r]{0}}
\end{picture}
\caption{Photon spectrum for different spin states
of muonic atom.}
\end{center}
\end{figure}

We are now ready to discuss the results of our work.
We start by looking at the role of the pion propagators.
The momentum transfer, $q_J$, is always space-like,
i.e., $q_J^2\simeq -\vec{q}_J^2$, 
due to the on-shellness of the incoming and outgoing nucleons.
This is the reason why $\Delta_\pi(q_J)$ is suppressed with
the important contribution coming from $f_1^V$ and $f_1^A$ 
instead of from $f_2^A$ and $f_3^A$  in 
Eqs.(\ref{f_form;f1v},\ref{f_form;f1a},\ref{f_form;f2a},\ref{f_form;f3a}).
On the other hand, $q_W^2$ increases almost linearly  with $E_\gamma$ 
and becomes time-like when  $E_\gamma$ is greater than $\sim 50$ MeV, 
since $q_W^{\ 2}\simeq 2m_\mu E_\gamma-m_\mu^2$.
Note that this is the region where the photon energy spectrum is established
in the experiment.
Hence  $\Delta_\pi(q_W)$ is enhanced in a high photon energy region
while  $\Delta_\pi(q_J)$ is always suppressed.
Consequently the LO distribution comes mainly from the
three terms $f_1^{V,LO}$, $f_1^{A,LO}$ and $A_{8}^{LO}$.
In particular, the contribution from  $A_{8}^{LO}$, 
the so-called Kroll-Ruderman (KR) term, 
carries about thirty five to sixty percents 
of the photon spectrum for 
$E_\gamma\ge 60 MeV$.

In Fig. 2 the spin averaged photon energy spectra for the LO, NLO and
${\rm N^2LO}$ contributions are plotted.
We find that the result of the phenomenological model
\cite{fearing1} can be more or less reproduced by the LO and 
NLO contributions.
For the experimentally measured region of the photon energy,
the NLO contribution remains  within 20 \% of the LO
contribution.

To summarize, we found that the next-order correction to the ${\rm NLO}$
description is negligible and does not remove the discrepancy present
at that order:
The further correction to the spectrum does not change 
appreciably the results of the previous
theoretical calculations\cite{fearing1,opat}.

This may seem disappointing in the sense that the persistent puzzle is
not resolved by our higher order calculation. On the other hand, our 
calculation is tightly under control and the fact that the next-order
terms to the ${\rm NLO}$ contribution are negligible implies that our 
theoretical treatment has converged. It is then legitimate to ask what
mechanisms other than strong-interaction dynamics could be the cause of
the discrepancy. 
As an illustration of such alternative mechanisms, we have considered the 
effects of various spin states in which the muonic atom
could be formed. The possible photon spectra for these
spin states are given in Figure 3. It is interesting to see that if
one assumed only the triplet state of the atom to be occupied, then
one would reproduce the  observed photon energy spectrum. While we are
not claiming that this could account for the discrepancy, such non-strong
interaction mechanisms could not be ruled out. 
Given the theoretical confidence in calculating higher-order chiral
corrections, it seems imperative that the presently available
experiment be  re-scrutinized
or that more refined measurements be made before concluding that the
constant $g_P$ is so drastically deviating from the Goldberger-Treiman
value.

\vskip 1cm

This work is supported in part by the Korea Science and Engineering
Foundation through Center for Theoretical Physics of Seoul National
University, and in part by Korea Ministry of Education (BSRI-97-2418). 
We are very grateful to Mannque Rho for his helpful  remarks and
suggestions, and to Harold Fearing for suggesting this problem.

\end{document}